# Bulletin of the Atomic Scientists

**Threats from space: 20 years of progress**
John L. Remo and Hans J. Haubold

# Threats from space: 20 years of progress

John L. Remo and Hans J. Haubold

## Abstract

It has been 20 years since planning began for the 1995 United Nations International Conference on Near-Earth Objects. The conference proceedings established the scientific basis for an international organizational framework to support research and collective actions to mitigate a potential near-Earth object (NEO) threat to the planet. Since that time, researchers have conducted telescope surveys that should, within the coming decade, answer many questions about the size, number, and Earth impact probability of these objects. Space explorations to asteroids and comets have been successfully carried out, including sample recovery. Laboratory experiments and computer simulations at Sandia National Laboratories have analyzed the effects of high-energy-density soft x-ray radiation on meteorites—which might help researchers develop a way to redirect an incoming asteroid by vaporizing a thin layer of its surface. An Action Team on NEOs, established in 2001 in response to recommendations of the Third United Nations Conference on the Exploration and Peaceful Uses of Outer Space, identified the primary components of NEO mitigation and emphasized the value of finding potentially hazardous NEOs as soon as possible. Recommendations from the action team are meant to ensure that all nations are aware of the NEO danger, and to coordinate mitigation activities among nations that could be affected by an impact, as well as those that might play an active role in any eventual deflection or disruption campaign.

On the morning of February 15, 2013, a meteor nearly 30 times brighter than the sun streaked into Earth's atmosphere at a speed of about 20 kilometers per second—almost 60 times the speed of sound—and exploded in the air near the Russian city of Chelyabinsk. A shock wave hit the ground, followed by meteorite fragments, causing many injuries and substantial damage but, fortunately, no fatalities (Popova et al., 2013). The object responsible for this sudden and dramatic event was an asteroid with an explosive energy equivalent to about a half-million tons of TNT. The Chelyabinsk impact and explosion is thought to be the most energetic since the poorly observed Tunguska explosion of 1908, which had an estimated energy of 3 to 15 megatons of TNT (Boslough and Crawford, 1997; Svetsov and Shuvalov, 2008).

A global survey of airbursts with energy equivalents of at least one kiloton of TNT has found dozens of these explosions since 2000, all thought to be caused by asteroid fragments entering Earth's atmosphere (Brown et al., 2013). A worry is that such an unanticipated explosion over a conflicted region might be misinterpreted as an act of war, with serious consequences.

Geological and paleontological records reveal that near-Earth objects (NEOs) of even larger size have struck Earth in the past, and astronomers say that it is practically inevitable that an object from outer space will threaten regional or even global civilization at some point in the future. The positive news is that, given sufficient warning time, the people of Earth have a good chance of mitigating this danger if they act collectively and decisively.

The organization best suited to coordinate NEO mitigation is the United Nations. That is why one of us called for and began organizing the United Nations International Conference on Near-Earth Objects in 1994 (Remo, 1996). It is now time to look back on what has been achieved, politically and scientifically, in the past 20 years—and to consider what can be expected in the future. Even after two decades of progress, there is still an urgent need to find, characterize, and develop plans for responding to potentially hazardous asteroids and comets—those that are large enough, and pass close enough to Earth's orbit, that they "might conceivably yield a globally-damaging impact in the foreseeable future" (Marsden, 1997).

## An international problem

Held at UN headquarters in New York in 1995, the United Nations conference established a scientific basis and an international organizational framework within which the NEO hazard to the inhabitants of Earth could be discussed and possibly mitigated. The publication of the conference proceedings outlined critical scientific and engineering topics that must be addressed to successfully meet the challenges of effective NEO mitigation—that is, the deflection or destruction of potentially hazardous asteroids and comets, with deflection as the preferred option because it is less risky.

During the past 20 years, there has been much to report on UN activities and research focused on near-Earth objects. Scientists have expanded NEO population surveys, launched space missions to asteroids and comets, done basic laboratory analytical research aimed at learning more about the composition of these objects, created computer models to study the potential for intercepting and diverting Earth-threatening objects, and made observations of atmospheric impact events. We cannot comprehensively describe, or even list, all of this work here. However, we can highlight some of the most important discoveries, political activities, and new challenges that have arisen since the publication of the conference proceedings (Remo, 1997) and lay the groundwork for future UN activities.

## Finding them before they find us

The first step in dealing with the NEO specter is to conduct space surveys aimed at identifying the most dangerous objects. NASA-sponsored surveys using ground-based telescopes dominate the field and are thought to have found most of the asteroids that are hundreds of meters or more in diameter, big enough to wipe out entire civilizations (Yeomans, 2013). Asteroids typically approach closer to Earth than comets, and are more numerous, so they are the focus of most observations—although a comet impact would be catastrophic to life on Earth.

Ground-based search programs supported by NASA include MIT's Lincoln Near-Earth Asteroid Research (LINEAR) telescope in New Mexico, the Catalina Sky Survey telescopes in Arizona and Australia, the Spacewatch project that uses telescopes on Arizona's Kitt Peak, and the Panoramic Survey Telescope and Rapid Response System (Pan-STARRS) in Hawaii. Using the space-based Wide-field Infrared Survey Explorer (WISE) telescope launched in December 2009, the NEOWISE project hunted for asteroids from September 2010 to February 2011 and was reactivated in December 2013 for a three-year mission focused on near-Earth objects. NEOWISE is the most accurate survey to date of the size distribution of NEOs, but there is still a huge range of uncertainty.

Despite the information provided by surveys, NASA's 2005 congressional directive to identify the orbits of 90 percent of the extremely dangerous NEOs —those that are at least 140 meters in diameter—by 2020 is behind schedule, with only about 10 percent found as of July 2013 (Irion, 2013). Furthermore, only 1 percent of the estimated million near-Earth asteroids have been identified. As of May 6, NASA had found 11,023 near-Earth objects, 862 of which are at least a kilometer wide, and 1,469 of which have been classified as potentially hazardous asteroids (NASA, 2014). On the positive side, NASA's program is discovering asteroids at a rapidly increasing rate.

Clearly, deploying new orbital telescopes would help to increase the discovery rate of NEOs even further. Infrared telescopes orbiting the sun near Venus are critical for this task but can only alleviate, not eliminate, the NEO risk—especially from smaller asteroid fragments like the one that exploded near Chelyabinsk (see Figure 1). Other space telescopes, orbiting the Earth, would help extend the search for these smaller-diameter objects (Mainzer et al., 2011). Russia plans to launch a new detection telescope in 2017 and also to extend the International Scientific Optical Network, which already includes 32 facilities in 13 countries (Russian Academy of Sciences, 2013). These new telescopes should provide a better picture of the NEO population—such as how the sizes of these objects compare with their numbers and their potential for Earth impacts—within the coming decade.

## Exploring outer space

Visits to asteroids and comets can tell astronomers even more about these objects than telescopic observation. Outstanding among several successful explorations was Japan's mission to the small near-Earth asteroid Itokawa. The unmanned Hayabusa spacecraft rendezvoused with Itokawa in 2005, studied its characteristics, landed on it, and returned to Earth in 2010 carrying samples with a composition matching that of LL chondrites, the most common meteorite type on Earth. Chemical, mineralogical, and isotopic analysis determined that the samples from Itokawa were altered by "space weathering" on the asteroid's surface over millions of years in orbit. These results established a relationship between telescope

**Figure 1.** Near-Earth asteroid fragment explodes over Chelyabinsk, Russia, in February 2013

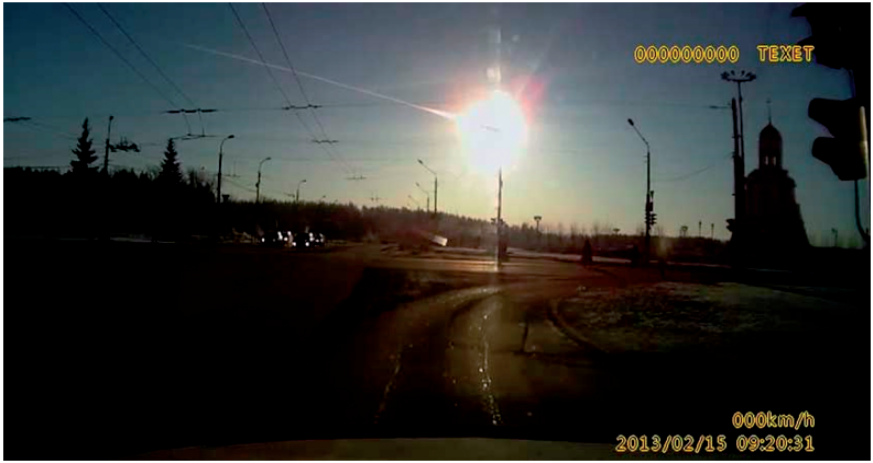

*Photo credit*: Взрыв метеорита над Челябинском

observations of asteroids and laboratory examinations of meteorites found on Earth.

There are different types of asteroids, and they may have variations in structure. For example, some may be more like flying rubble piles than solid masses; some may have massive cracking; and some may be more porous than others. Understanding these characteristics is critically important for estimating what might happen if Earthlings tried to redirect a dangerous asteroid.

NASA's Dawn spacecraft may help answer some questions about the composition and origins of asteroids. Launched in 2007, Dawn has already visited the asteroid Vesta and is expected to arrive at the asteroid Ceres in February 2015. These are the two most massive asteroids known, but they are very different: Vesta is dry and rocky, while the ice-covered Ceres might have a liquid ocean beneath its surface. Close inspections of both will help astronomers understand these differences.

Variations in composition and structure are not the only challenge for an asteroid-diverting mission. Some near-Earth asteroids have very high rotation rates, or even ring-like orbital fragments, that could make it difficult to predict the effects of a collision with an interceptor sent from Earth. Heading off a comet bound for Earth would be even more difficult, because of uncertainties about comets' population distribution, size, varying compositions and structures, highly active evolution, and nongravitational interactions—for example, jetting emissions can affect their orbit.

The important and impressive Deep Impact mission, which arrived at the comet Tempel 1 in 2005, probed the structure and composition of the comet's 6-kilometer-wide nucleus by firing a massive copper impactor at it (A'Hearn et al., 2005). The collision excavated a crater and ejected a complex array of icy and dusty materials. The impact appeared to change the comet's orbit by only the order of 10 to 20 meters per year, which would not be enough to spare Earth from an incoming comet. But this is no surprise, and only underscores the need for high-energy payloads that would have a bigger impact (Remo et al., 2013). The Deep Impact mission showed that it was possible to rendezvous with a comet and change its momentum. The Stardust spacecraft revisited Tempel 1 in 2011 and observed what is believed to be a shallow crater carved out by the impactor; it is about 50 meters wide and surrounded by brighter material ejected during the impact. The Rosetta spacecraft, launched in 2004 by the European Space Agency, is anticipated to carry out landing and sampling on Comet Churyumov-Gerasimenko in August 2014, providing further insights into the composition of comets.

## Developing a defense

The most menacing near-Earth objects tend to be the biggest ones, and it will take a huge amount of energy to divert a massive object that is on a collision course with Earth. Since the 1995 United Nations NEO conference, there has been very little innovative computer modeling of how such an energy transfer might be accomplished. The possibilities include high-speed mechanical projectiles fired at the NEO, conventional explosives detonated at or below the surface of the NEO, and x-ray or neutron radiation such as that which might be produced by a nuclear explosive device (Shafer et al., 1997).

A high-speed projectile, for example, could shift an NEO's orbit by mechanically transferring momentum to the NEO, and would be most useful for deflecting relatively small NEOs with long warning times. For shorter warning times and larger NEOs, high-density radiation could be used to vaporize a thin layer of the NEO's surface, creating a jet of material that imparts a momentum change to the NEO. In either case, repeated interceptions may be necessary to move the potentially hazardous object sufficiently out of harm's way.

A long-term experimental study using the Z-pinch facilities at Sandia National Laboratories—also used to test nuclear fusion dynamics—simulated the effects of high-energy-density soft x-ray radiation on meteorite targets, which were used as surrogates for near-Earth asteroids (Remo et al., 2013). This study showed that radiation is a viable option for the orbital deflection of large objects with shorter warning times.

Simulation of high-energy mechanical and radiation interactions with NEO models is critical for designing a successful mitigation mission. Current rocket technology with minor modifications —for example, using an Ares I, Atlas V, Athena, or Delta IV launch vehicle—is thought to be up to the task of delivering a single high-energy payload—one mechanical impactor, for example—within a time frame when the NEO is most susceptible to a hit that would change its course. To deal with more massive NEOs within a short warning time, the Ares V system is capable of launching six independently targeted payloads. The magnitude of the hazard and the available time frame will determine the type and size of the required launch payload.

The 2013 Chelyabinsk explosion demonstrated the potential effects of near-Earth objects, as well as the difficulty of detecting them. The impact was well recorded because of the ubiquity of consumer video devices. Images of the event, along with timely and well-managed damage assessments, provided quantitative estimates of key parameters: a shallow entry angle of about 18 degrees, a speed of about 20 kilometers per second, an airburst altitude of about 27 kilometers, and a debris path of about 100 kilometers. Mineralogical analysis of recovered meteorite fragments indicate a strongly shocked LL chondrite with a density of 3.3 grams per cubic centimeter (similar to that of sand) and an entry body about 20 meters in diameter that weighed about 13,000 metric tons (roughly the size of a bus).

That the Chelyabinsk impact occurred without warning should come as no surprise, according to work by Syuzo Isobe and Makoto Yoshikawa (1997) presented at the 1995 conference, which showed that ground-based telescopes have a blind spot surrounding the sun in which an asteroid approaching the Earth cannot be seen because of the bright sky background. The number of NEOs coming from this area to within less than 0.01 astronomical units from the Earth (that is, less than one-hundredth the distance from the Earth to the Sun) is more than 30 percent of the total. One way to detect these NEOs is to deploy infrared-sensitive survey telescopes between the Sun and Venus, as discussed above. Syuzo Isobe (Isobe and Hirayama, 1998) organized an International Astronomical Union NEO conference, as a follow-up to the UN conference, in Kyoto, Japan, in 1997.

## An international action team

Since the 1995 UN conference, there have been numerous meetings, workshops, and topical conferences sponsored by member states of the UN Committee on the Peaceful Uses of Outer Space (COPUOS) to recommend, initiate, or develop scientific initiatives and topics laid out in 1995 (Isobe and Hirayama, 1998; Remo and Haubold, 2001; Wamsteker et al., 2004).

In 2001, COPUOS established an Action Team on Near-Earth Objects, in response to recommendations of the Third United Nations Conference on the Exploration and Peaceful Uses of Outer Space (Remo and Haubold, 2001; UNISPACE III, 1999) held in Vienna in 1999, and charged the team with reviewing the content, structure, and organization of ongoing efforts in the field of near-Earth objects; identifying any gaps in the ongoing work where additional coordination is required and/or where other countries or organizations could make contributions; and proposing steps for the improvement of international coordination in collaboration with specialized bodies. Since 2001, the action team has considered annual reports submitted by the member states active in NEO work, as well as recommendations concerning an international response made at workshops and conferences conducted by the international NEO community.

The action team identified three primary components of mitigation: discovering hazardous asteroids and comets and identifying those objects requiring action; planning a mitigation campaign that includes deflection and/or disruption actions and civil defense activities; and implementing a campaign, if the threat warrants. The action team emphasized the value of finding hazardous NEOs as soon as possible in order to avoid unnecessary delays in planning and launching mitigation missions. The team called for establishing an international asteroid warning network that would link existing institutions and recommended that space agencies establish a space mission planning advisory group to consider options for planetary defense. Recommendations by the action team are meant to ensure that all nations are aware of potential hazards, and to ensure the design and coordination of mitigation activities among nations that could be affected by an impact and those that might play an active role in any eventual deflection or disruption campaign (COPUOS, 2013).

## A double-edged sword

At its 68th session on December 11, 2013, the General Assembly of the United Nations embraced the recommendations for an international response to the near-Earth object impact threat, which were endorsed by the Scientific and Technical Subcommittee of COPUOS and by COPUOS (United Nations General Assembly, 2013). As the COPUOS action team noted in its recommendations, "The risk that an asteroid would impact the Earth is extremely small, but, depending on its size and impact point, the consequences could be catastrophic. However, perhaps uniquely among natural hazards, there is the potential to prevent near-Earth object (NEO) impact events through timely actions" (COPUOS, 2013). The emphasis in the coming decades should be on mitigating the potential impacts of large NEOs, rather than relatively small ones that can be consumed by the Earth's atmosphere—such as the asteroid fragment that exploded near Chelyabinsk.

There is also potential to prevent a colossal misunderstanding. NEOs pose a dual danger to civilization: not just the possibility of a catastrophic collision with a massive object, but also the chance that a near-Earth object entering the atmosphere might be mistaken for an intentional attack. The effort to prevent a celestial object from causing massive extinction of life on Earth is a project that can unite nations around the world in the peaceful use of outer space. Given the long lead time needed to develop technology that can safely and reliably divert a threatening object, now is the time for an open discussion of means and methods.

## Funding

This research received no specific grant from any funding agency in the public, commercial, or not-for-profit sectors.

## Author biographies

Physicist **John L. Remo** is a research associate at the Harvard-Smithsonian Center for Astrophysics and the Departments of Astronomy and Earth and Planetary Sciences at Harvard University, USA. His expertise is in planetary science, astrophysics, and laser physics. He organized the 1995 United Nations International Conference on Near-Earth Objects. He used the Z facilities at Sandia National Laboratories to simulate the effects of high-energy-density soft x-radiation on planetary bodies; to simulate the origin of solar system–like planets; to explore the structure of exoplanets; and to study the potential for deflecting near-Earth objects in the future. He received the Nininger Meteorite Award for his prediction and discovery of the ductile-brittle transition in meteoritic irons. Minor planet Remo 2114 T-2, also known as asteroid 9137 Remo, is named after him in recognition of his work in laser physics and planetary science; it is not a potentially hazardous object.

Astrophysicist **Hans J. Haubold** is a professor of theoretical astrophysics at the United Nations Office for Outer Space Affairs in Vienna and the Centre for Mathematical Sciences in Pala, India. His expertise is in solar physics and neutrino astrophysics. From 1991 to 2012, he organized the UN Basic Space Science Initiative (Haubold, 2013) with support from the European Space Agency, NASA, the Japan Aerospace Exploration Agency, and member states of the United Nations. The issue of near-Earth objects was part of this initiative and was introduced to the United Nations through the UN International Conference on NEOs.